# A sub-1-volt analog metal oxide memristive-based synaptic device for energy-efficient spike-based computing systems

Cheng-Chih Hsieh[1], Anupam Roy[1], Yao-Feng Chang[1], Davood Shahrjerdi[2], and Sanjay K. Banerjee[1]

A memristor is a two-terminal 'memory resistor' electronic device, in which a metal oxide switching layer is sandwiched between two metal electrodes [1-4]. In general, memristors offer non-linear switching characteristics, and materials and process compatibility with advanced silicon manufacturing. These attributes have spurred the exploration of memristors as synaptic devices for realizing spike-based hardware learning systems that are capable of processing unstructured, temporal data [5-10]. However, for memristor-based technologies to be viable, the device should exhibit several key characteristics. It should have a compact nanoscale footprint, operate at a voltage close to 1V that is compatible with complementary metal oxide semiconductor (CMOS) technology, have reproducible electrical characteristics, and possess high switching speed to minimize the energy consumption [11]. Furthermore, the hardware integration of synaptic connections in advanced neural networks requires memristors with multiple resistive states [12, 13]. These are challenging requirements and are difficult to implement without significant innovations.

The phenomenological principle of memristor device operation is based on the change in the physical properties of a conductive filament (associated with the presence of oxygen vacancies) by applying an electric field across the metal oxide switching layer [14-16]. The resulting motion of the oxygen vacancies alters the device resistance between low (Set) and high (Reset) states, depending on the direction and the amplitude of the electric field. So far, a variety of structures from a large set of materials (various metal oxide switching layers and metal electrodes) have been studied in the literature [4, 17, 18]. Several key findings can be drawn from those studies regarding the performance, energy and scalability of this type of devices. The most important finding reveals the trade-off between the switching energy and the data retention time—that is often referred to as voltage-time dilemma [19]. This trade-off is associated with the energy barrier of the device structure. For example, devices made of metal oxides with small energy bandgap ($E_g$), such as titanium oxide ($TiO_x$, $E_g$~3.4eV), generally exhibit low operating voltage and compromised data retention [20], while those with large bandgap, such as hafnium oxide ($HfO_x$, $E_g$~5.4eV) demonstrate the opposite [21]. However, the fabrication of devices with bilayer switching stacks has shown to be effective in mitigating this trade-off. In particular, the improvement in data retention was obtained by the incorporation of an ultra-thin wide bandgap metal oxide capping layer (for example aluminum oxide) [22]. On the other hand, the addition of a reactive capping metal (for example titanium, hafnium, etc.) as an oxygen scavenging layer provided a pathway for reducing the operating voltage of the devices [23, 24]. Despite significant advances, a sub-1V memristive device that simultaneously affords built-in analog behavior, energy efficiency on par with a biological synapse, forming-free operation and low device-to-device variations is still elusive.

---

[1] Microelectronics Research Center, The University of Texas at Austin, Austin, Texas 78758, USA.

[2] Electrical and Computer Engineering, New York University, Brooklyn, New York, 11201, USA.

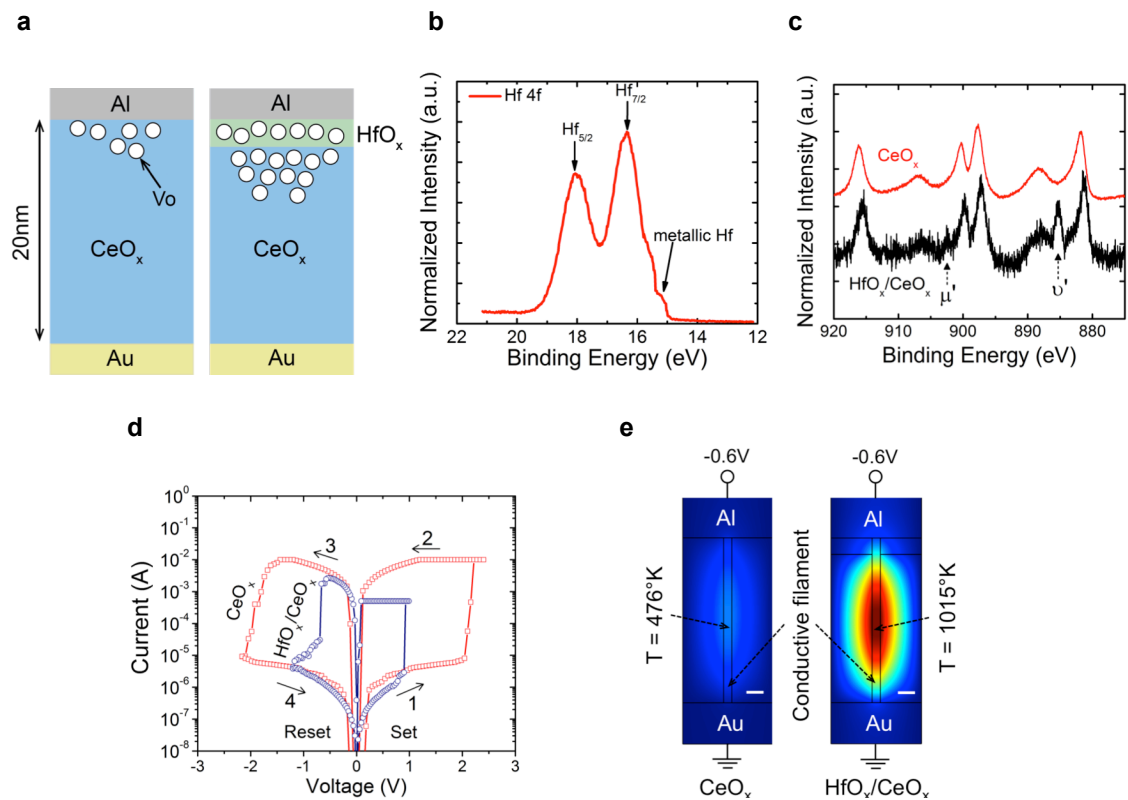

**Figure 1: Improving memristor device characteristics using an engineered sub-stoichiometric HfOx capping layer. a**, Schematic structure of two memristors with and without the engineered HfOx, conceptually illustrating the increase of the oxygen vacancy density in the CeOx switching layer. This attribute of the bilayer memristor results in the forming-free operation and the reduction of the Set voltage. XPS spectra of the **b**, engineered HfOx, and **c,** CeOx films with and without the HfOx capping layer. The XPS studies indicate the increase of the oxygen vacancy concentration in the CeOx film capped with the oxygen-deficient HfOx layer. **d,** Representative current-voltage characteristics of two memristors, indicating the sub-1V operation of the bilayer memristive device. **e**, Heat transfer simulations illustrate enhanced Joule heating in the bilayer structure, causing the marked reduction of the Reset voltage (Scale bars are 2nm). The observed increase in Joule heating arises from the high thermal resistivity of HfOx at nanoscale.

Here, we have developed a memristive-based synaptic device by engineering the material properties of an HfOx capping layer in a bilayer structure with a cerium oxide (CeOx) switching layer. In this structure, the combination of sub-stoichiometric structural properties of the HfOx capping layer and its enhanced thermal resistivity at nanoscale dimensions leads to the significant improvement in switching behavior of the devices in terms of the operating voltages, device performance uniformity, reproducibility, and reliability. Furthermore, this structure yields forming-free devices with an analog resistance state that is inherent to the device itself. This key attribute of our HfOx/CeOx devices enables the implementation of Hebbian learning [25], validating the plasticity of the synaptic connection.

Memristor devices (150µm diameter) were fabricated on silicon substrates capped with 300nm silicon dioxide. The device structure consists of gold bottom electrode, $HfO_x/CeO_x$ switching layer, and aluminum top electrode. The total thickness of the bilayer switching layer in all experiments was kept at 20nm, while varying the thickness of the $HfO_x$ and $CeO_x$ layers. The metal electrodes were formed using electron-beam evaporation. Figure 1(a) conceptually illustrates the effect of the engineered $HfO_x$ capping layer on the concentration of oxygen vacancies in the $CeO_x$ switching layer. X-ray photoelectron spectroscopy (XPS) was performed to guide the development of the bilayer structure (See Supplementary Information). Figure 1(b) shows the Hf 4f spectrum of the engineered $HfO_x$ capping layer, revealing the sub-stoichiometric nature of the film. The data indicates the presence of Hf $4f_{7/2}$ and Hf $4f_{5/2}$ peaks at 16.32eV and 18.03eV, respectively—which is consistent with the previous reports in the literature [26, 27]. Metallic Hf was also found in the engineered $HfO_x$ layer, evident from the peak at 15.02 eV. The chemical composition of the $HfO_x$ was quantified using the Casa XPS software, in which x was found to be about 1.75.

Figure 1(c) shows the Ce 3d XPS spectra of the $CeO_x$ switching layer with and without the engineered $HfO_x$ capping layer. In these experiments, the $CeO_x$ and $HfO_x$ layers were 20nm and 0.8nm, respectively. As a result, the adequately small thickness of the $HfO_x$ capping layer allowed the XPS analyzer to receive signal from the $CeO_x$ layer. As can be seen in Figure 1(c), the bilayer structure exhibits discernable u' and v' peaks at 904 eV and 885 eV [28-30] that are absent in the spectrum of the $CeO_x$ layer with no $HfO_x$ capping layer. The u' and v' peaks signals the reduction of the $Ce^{4+}$ to $Ce^{3+}$ states [29, 30], which can be translated to the formation of excess oxygen vacancies at regions near the $HfO_x/CeO_x$ interface. The marked increase of the oxygen vacancy concentration in the bilayer structure permits the formation of the conductive filament using a smaller electric field, thereby enabling the low-voltage operation of the bilayer structure. Figure 1(d) shows the representative dc current-voltage characteristics of two $CeO_x$-based devices with and without the engineered $HfO_x$ capping layer, demonstrating significant reduction of the Set voltage to below 0.8V.

In a memristive device, the transition from low to high resistance states occurs as the polarity of the electric field across the device is reversed. As the reverse electric field increases, the oxygen anions in the conductive filament begin to disperse through drift and diffusion processes [10]. Considering the similar thickness of the switching layer in Figure 1(d), the improved Reset voltage of the bilayer device may be explained by locally enhanced diffusion of oxygen vacancies. We infer that the enhanced thermal resistivity of $HfO_x$ at nanoscale dimensions amplifies Joule heating in the $CeO_x$ switching layer, thereby accelerating the dispersion of oxygen anions at a lower electric field. To elucidate this concept, we performed numerical heat transfer analysis using COMSOL simulator for two devices in Figure 1(e) at the bias of -0.6V. The simulation results indicate significant enhancement of Joule heating in the bilayer structure. For these simulations, we used the measured electrical parameters of the layers, while the thermal parameters were obtained from the literature [28, 31-35].

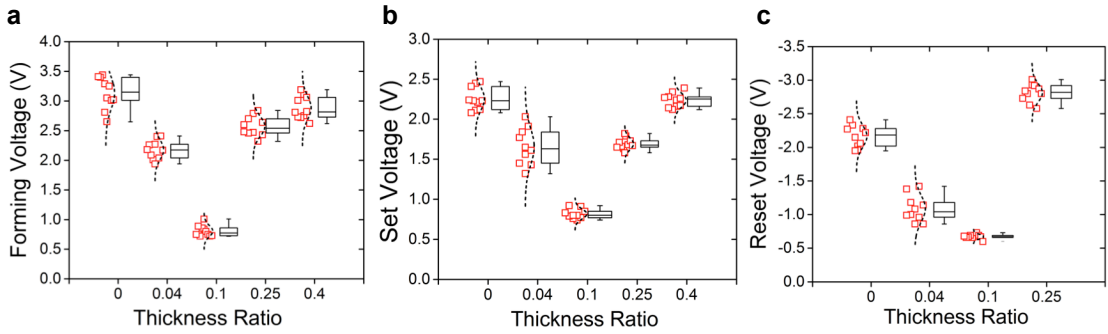

**Figure 2: Effect of HfO$_x$/CeO$_x$ thickness ratio on the memristor device behavior.** The data indicates that the optimal device characteristics (**a**, forming voltage, **b**, Set voltage, and **c**, Reset voltage) occurs at the thickness ratio of about 0.1. Moreover, the device-to-device variation is reduced at this optimal thickness ratio. The equivalency of the forming and Set voltages at the optimal thickness ratio confirms the forming-free operation of the device.

Low device variability is critical for implementation of large neural networks with high density of memristive synaptic connections. Therefore, we statistically examined the effect of the HfO$_x$ thickness on the important device parameters: Set, Reset, and forming voltages. In these experiments, the HfO$_x$/CeO$_x$ thickness ratio was varied, while keeping the total thickness of the bilayer stack fixed at 20nm. The data in Figure 2 indicates that the insertion of an HfO$_x$ capping layer with the optimal thickness ratio of about 0.1 significantly improves the uniformity of the key device parameters. Interestingly, this optimal thickness ratio also coincides with the minimum operating voltages of the bilayer structure. We surmise that the HfO$_x$ film begins to act as an independent switching layer beyond this optimal thickness ratio, resulting in significant increase in both the device operating voltages and the device variability. Moreover, the Reset voltage begins to increase as the HfO$_x$ film becomes thicker. This observation is in agreement with our heat transfer simulation results in Figure 3.

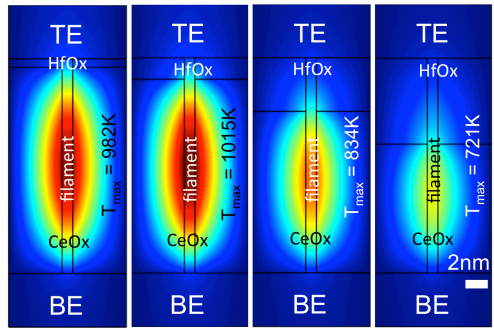

**Figure 3: Effect of HfO$_x$ film thickness on Joule heating.** Numerical heat transfer simulation results for several bilayer HfO$_x$/CeO$_x$ structures with varying HfO$_x$ to CeO$_x$ thickness ratio at the bias voltage of -0.6V. The total thickness of the HfO$_x$/CeO$_x$ stack was kept at 20nm. The Joule heating begins to diminish as the thickness of the HfO$_x$ was increased, which arises from the thickness dependence of the HfO$_x$ thermal conductivity [35].

A fresh memristive device generally requires an initial electroforming step; that is the formation of a conductive filament using a relatively large electric potential (known as the forming voltage) before the device can operate at normal Set and Reset voltages. The disparity between the Set and the forming voltages necessitates that the devices in a crossbar array are isolated and accessed individually for electroforming [36, 37] in order to avoid the breakdown of the neighboring 'formed' devices in the array. However, the physical constraints of these strategies limit the implementation of high-density crossbar arrays. Our bilayer $HfO_x/CeO_x$ device is free from such a limitation, exhibiting forming-free behavior; i.e. the Set voltage is adequate to form the conductive filament in a fresh memristive device (See Figure 2(a) and (b)). This characteristic is attributed to the efficacy of the $HfO_x$ capping layer in creating sufficiently high concentration of excess oxygen vacancies in the $CeO_x$ switching layer.

Figure 3 shows the heat transfer simulations for devices with varying $HfO_x/CeO_x$ thickness ratio, for which the total thickness of the $HfO_x/CeO_x$ stack was 20 nm. The peak temperature value was found to be the highest when the thickness ratio was about 0.1. The simulation results suggest that capping the $CeO_x$ switching layer with a sufficiently thin layer of $HfO_x$ enhances the Joule heating, owing to the pronounced thermal resistivity of $HfO_x$ at nanoscale. However, as the thickness of the $HfO_x$ increases, the Joule heating begins to diminish, which is consistent with the thickness dependence of the $HfO_x$ thermal conductivity [35]. The enhanced Joule heating effect in the optimal structure is therefore expected to enhance the diffusion of the oxygen vacancies during the Reset process, thereby reducing the Reset voltage.

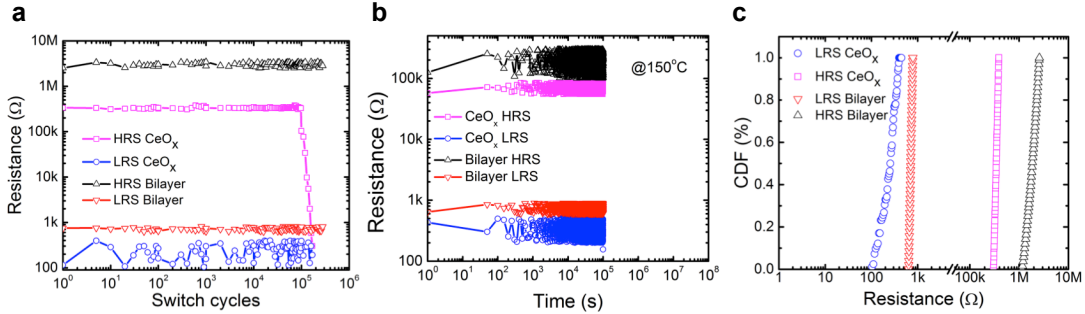

**Figure 4: Device reliability studies. a,** The endurance test results for the $CeO_x$ and the optimal $HfO_x/CeO_x$ devices. In addition to the improved endurance properties, the bilayer device exhibits larger HRS and LRS values compared to the device with no $HfO_x$. The increase of the LRS and HRS values is favorable for reducing the switching power consumption of the bilayer device. **b,** The accelerated retention test for the $CeO_x$ and the $HfO_x/CeO_x$ devices measured at 150°C at constant stress voltage of +0.2V. The results indicate projected data retention of 10 years for both devices. **c,** Representative CDF plot of the cycle-to-cycle programing characteristics for two devices with and without the engineered $HfO_x$ layer.

The bilayer structure exhibits excellent switching reliability at the thickness ratio of 0.1, which conceivably stems from the reduced operating voltage of the device. In Figure 4(a), the optimal memristor bilayer structure survives more than $2 \times 10^5$ cycles of programing (endurance test). The accelerated retention test in Figure 4(b) indicates projected data retention of 10 years for the bilayer devices (See Supplementary Information). The cumulative distribution function (CDF) in Figure 4(c) illustrates the representative switching characteristics between different programing cycles for two $CeO_x$ devices with and without the engineered $HfO_x$ capping layer. The CDF plot indicates the improved uniformity of the On-state performance between programming cycles of the same bilayer device, while the Off-state characteristic of the device appears to have been degraded, perhaps due to the non-uniformity of the Joule heating effect. The bilayer device exhibits average low- and high-resistance states (LRS and HRS) of about 600Ω and 2.8MΩ that are larger than those of the device with no HfO$_x$ capping layer by factors of 4 and 10, respectively. The resulting increase in the LRS and HRS values is beneficial for reducing the switching power consumption of the device during the Set and Reset operations.

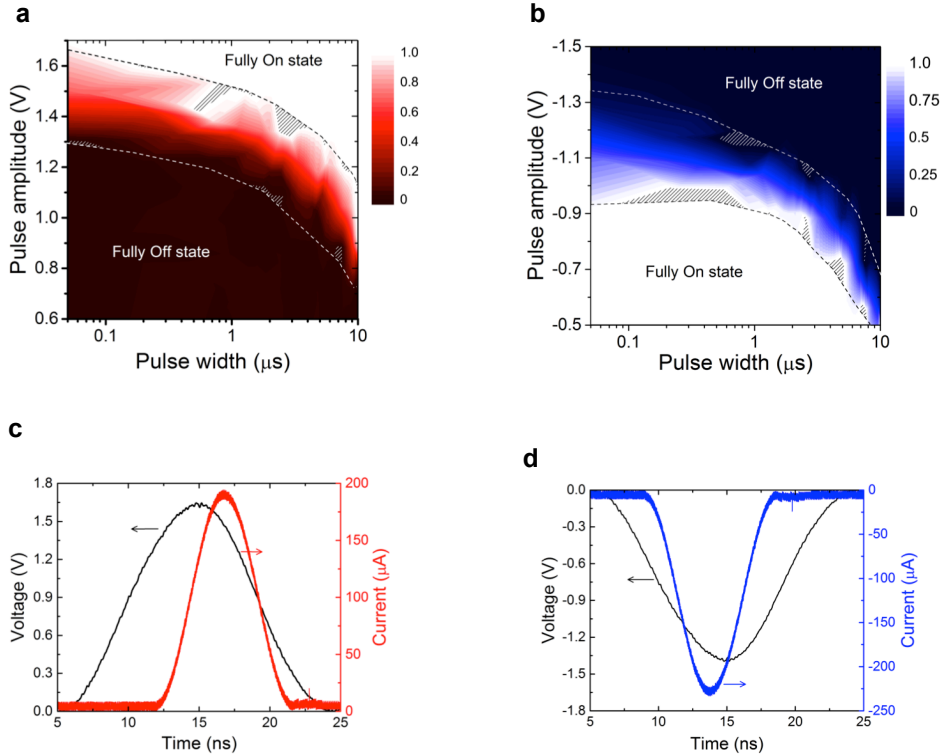

**Figure 5: Analog memory characteristic of the bilayer memristor.** The normalized conductance of a bilayer memristor is plotted as a function of pulse widths and amplitudes when the device switches from **a**, fully Off state to fully On state, and **b**, fully On state to fully Off state. The dashed lines are guide to the eye and the hatched regions denote unmeasured points. The data in **a**, and **b** reveal the gradual change in the conductance of the device between the fully Off and On states. Full On/Off switching energy consumptions of ~2.6 and 2.1pJ were calculated from the transient **c**, Set and **d**, Reset voltage and current waveforms, respectively.

The series connection of one transistor and one memristor (1T-1R) is a popular approach for implementing multi-state memory function [38]. The use of such configurations, however, limits the memristor integration density because of the physical constraints imposed by the transistor dimensions as well as the need for a complicated driver circuit in order to independently control each transistor. To circumvent these practical issues, the multi-state characteristic must be inherent to the two-terminal memristive device itself. Figures 5(a) and 5(b) illustrate the pulse measurement results for a bilayer device (with the optimal structure), indicating the gradual change in the conductance of the filament between the fully On and Off states. The observed resistive states are inherent to the device because no current compliance limit was used during these measurements.

Interestingly, the bilayer device also exhibits weak voltage-time dependence for pulses shorter than a few microseconds, which could be attributed to the dominant effect of the $HfO_x$ capping layer on the device switching behavior. The full On/Off energy consumption during Set and Reset steps was calculated to be, respectively, 2.6 and 2.1 pJ using the corresponding transient voltage and current waveforms in Fig. 5(c) and 5(d). Considering the analog characteristic of the resistive states together with the large HRS to LRS ratio in excess of $10^3$, the energy consumption for switching between the intermediate resistance states will be much smaller (about tens of fJ, assuming memory states with an increment of $100\Omega$).

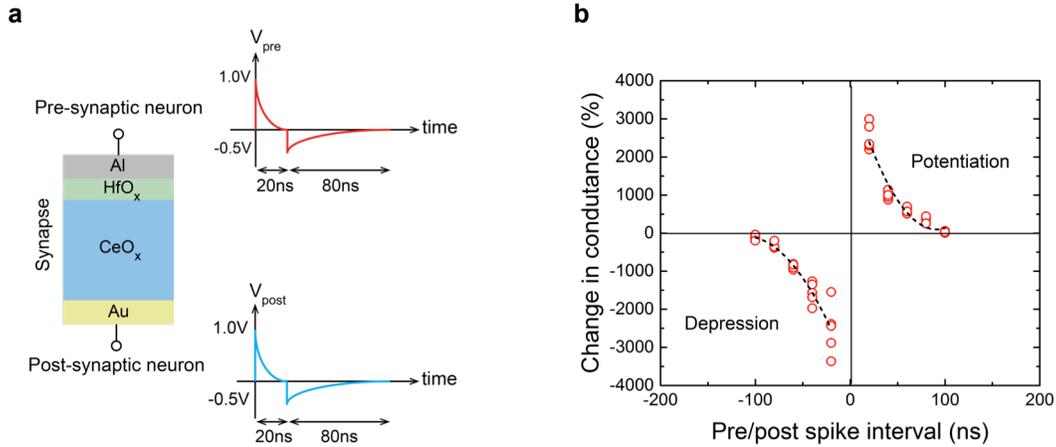

**Figure 6: Implementation of STDP learning using the $HfO_x/CeO_x$ memristive device. a**, Schematic representation of the learning experiment. Two waveforms with identical shapes were applied to the top and bottom electrodes. In the learning experiments, the time intervals between the pre- and post-synaptic spikes were varied in order to probe the synaptic depression ($\Delta t<0$) and potentiation ($\Delta t>0$). The positive (negative) time difference indicates that the pre-synaptic spike occurs before (after) the post-synaptic one. **b**, The plot clearly indicates the marked change in the synaptic strength as a function of different pre/post spike intervals.

Inspired by the brain, spike-based hardware learning systems have potential to be efficient and compact for processing unstructured data [39]. In such systems, the learning mechanism follows the spike-based form of Hebbian learning [25], i.e. STDP, in which the change in the strength of the synapse depends on the time difference ($\Delta t$) between the pre- and post-synaptic neural spikes. Figure 6(a) illustrates the synaptic waveforms. The waveforms with exponential decays were emulated with a series of square pulses (Supplementary Fig. S1). For these experiments, we have chosen an average spike rate of about 1MHz, which is $10^5$ times faster than that of the brain. This corresponds to a time step of ~1µs for updating the internal state of neurons and calculating the synaptic currents, assuming the neuron spiking probability of 0.01 as in the brain. Note that the acceleration of the learning rate is beneficial for handling large amount of data, while allowing the reduction of the energy consumption of the memristors. Figure 6(b) shows the plot of the normalized conductance change of the optimal bilayer device as a function of the time difference between the pre- and post-synaptic neural spikes. The data is fitted with exponential decay functions, confirming an STDP behavior similar to that of a biological synapse. Moreover, the data indicates a remarkable change in the normalized conductance of the device (>30 times) when the pre- and post-synaptic spikes overlap.

In summary, we have demonstrated a new bilayer $HfO_x/CeO_x$ memristors by tailoring the structural properties of the nanoscale $HfO_x$ capping layer. The memristive device was readily implemented using CMOS-compatible materials and processes. The device is forming-free and thus amenable to high-density integration. More importantly, this device also exhibits analog resistance states, sub-1V operating voltages, and energy efficient operation. Furthermore, the STDP learning rule was successfully implemented, following the Hebb's rule of learning; that is, neurons fire together wire together. The salient features of this new memristor meet the main requirements for a native synaptic device and can be used for hardware implementation of STDP-based learning systems.

## Acknowledgement

The authors acknowledge the financial support by the NSF NNCI program, and the NASCENT NSF ERC grant.

## Methods

The $CeO_x$ layer was reactively evaporated in oxygen plasma ambient at 0.2mTorr and an average deposition rate of ~0.06nm/s. The $HfO_x$ layer was formed by plasma-assisted atomic layer deposition (PE-ALD) using water and tetrakis (dimethylamido) hafnium ($Hf(NMe_2)_4$) precursors. The film optimization involved varying a wide range of deposition conditions. The optimal $HfO_x$ capping layer was deposited at 200°C. The pulse width of the hafnium precursor was 0.25s and the hold time between each pulse was 5s. The optimal oxygen plasma power was found to be 300W. The devices were isolated using a wet etching process by first patterning the $HfO_x$ film in buffered oxide etch followed by removing the $CeO_x$ layer in a mixture of hydrochloric acid, potassium hexacyanoferrate, and de-ionized water. Devices were measured under vacuum in a Lakeshore CRX-VF Probe Station using Agilent semiconductor parameter analyzer B1500 equipped with a Semiconductor Pulse Generator Unit (SPGU). Care was taken to minimize the impact of parasitic elements, for example, capacitances.

## Author contributions

C-C.H., D.S., and S.K.B conceived the experiments. C-C.H. performed device fabrication, electrical and material characterizations, and COMSOL simulations. A.R. contributed to XPS characterizations, Y-F.C. contributed to device fabrication and electrical characterizations. All authors discussed the results. D.S., C-C.H., and S.K.B co-wrote the paper.

## Additional information

The authors declare no competing financial interests. Correspondences and request for materials should be addressed to C-C.H.

# References

1.      Chua LO. Memristor-the missing circuit element. Circuit Theory, IEEE Transactions on. 1971;18(5):507-19.

2.      Strukov DB, Snider GS, Stewart DR, Williams RS. The missing memristor found. nature. 2008;453(7191):80-3.

3.      Waser R, Aono M. Nanoionics-based resistive switching memories. Nature materials. 2007;6(11):833-40.

4.      Wong H-SP, Salahuddin S. Memory leads the way to better computing. Nature nanotechnology. 2015;10(3):191-4.

5.      Kuzum D, Yu S, Wong HP. Synaptic electronics: materials, devices and applications. Nanotechnology. 2013;24(38):382001.

6.      Masquelier T, Guyonneau R, Thorpe SJ. Competitive STDP-based spike pattern learning. Neural computation. 2009;21(5):1259-76.

7.      Yu S, Wu Y, Jeyasingh R, Kuzum D, Wong H-SP. An electronic synapse device based on metal oxide resistive switching memory for neuromorphic computation. Electron Devices, IEEE Transactions on. 2011;58(8):2729-37.

8.      Jo SH, Chang T, Ebong I, Bhadviya BB, Mazumder P, Lu W. Nanoscale memristor device as synapse in neuromorphic systems. Nano letters. 2010;10(4):1297-301.

9.      Prezioso M, Merrikh-Bayat F, Hoskins B, Adam G, Likharev KK, Strukov DB. Training and operation of an integrated neuromorphic network based on metal-oxide memristors. Nature. 2015;521(7550):61-4.

10.     Yang JJ, Strukov DB, Stewart DR. Memristive devices for computing. Nature nanotechnology. 2013;8(1):13-24.

11.     Kuzum D, Jeyasingh RG, Yu S, Wong H-SP. Low-energy robust neuromorphic computation using synaptic devices. Electron Devices, IEEE Transactions on. 2012;59(12):3489-94.

12.     Indiveri G, Legenstein R, Deligeorgis G, Prodromakis T. Integration of nanoscale memristor synapses in neuromorphic computing architectures. Nanotechnology. 2013;24(38):384010.

13.     Yu S, Kuzum D, Wong H-SP, editors. Design considerations of synaptic device for neuromorphic computing. Circuits and Systems (ISCAS), 2014 IEEE International Symposium on; 2014: IEEE.

14.     Yang JJ, Pickett MD, Li X, Ohlberg DA, Stewart DR, Williams RS. Memristive switching mechanism for metal/oxide/metal nanodevices. Nature nanotechnology. 2008;3(7):429-33.

15.     Kamiya K, Yang MY, Park S-G, Magyari-Köpe B, Nishi Y, Niwa M, et al. ON-OFF switching mechanism of resistive–random–access–memories based on the formation and disruption of oxygen vacancy conducting channels. Applied Physics Letters. 2012;100(7):073502.

16.     Hsieh C-C, Roy A, Rai A, Chang Y-F, Banerjee SK. Characteristics and mechanism study of cerium oxide based random access memories. Applied Physics Letters. 2015;106(17):173108.

17.     Akinaga H, Shima H. Resistive random access memory (ReRAM) based on metal oxides. Proceedings of the IEEE. 2010;98(12):2237-51.

18.     Jeong DS, Thomas R, Katiyar R, Scott J, Kohlstedt H, Petraru A, et al. Emerging memories: resistive switching mechanisms and current status. Reports on Progress in Physics. 2012;75(7):076502.

19.     Schindler C, Staikov G, Waser R. Electrode kinetics of Cu-SiO2-based resistive switching cells: Overcoming the voltage-time dilemma of electrochemical metallization memories. Applied physics letters. 2009;94(7):2109.

20.     Hu C, McDaniel MD, Posadas A, Demkov AA, Ekerdt JG, Yu ET. Highly controllable and stable quantized conductance and resistive switching mechanism in single-crystal TiO2 resistive memory on silicon. Nano letters. 2014;14(8):4360-7.

21.     Yu S, Chen H-Y, Gao B, Kang J, Wong H-SP. HfOx-based vertical resistive switching random access memory suitable for bit-cost-effective three-dimensional cross-point architecture. ACS nano. 2013;7(3):2320-5.

22.     Goux L, Fantini A, Kar G, Chen Y-Y, Jossart N, Degraeve R, et al., editors. Ultralow sub-500nA operating current high-performance TiNAl 2 O 3 HfO 2 HfTiN bipolar RRAM achieved through understanding-based stack-engineering. VLSI Technology (VLSIT), 2012 Symposium on; 2012: IEEE.

23.     Chen YY, Goux L, Clima S, Govoreanu B, Degraeve R, Kar GS, et al. Endurance/Retention trade-off on cap 1T1R bipolar RRAM. Electron Devices, IEEE Transactions on. 2013;60(3):1114-21.

24.     Goux L, Fantini A, Redolfi A, Chen C, Shi F, Degraeve R, et al., editors. Role of the Ta scavenger electrode in the excellent switching control and reliability of a scalable low-current operated TiNTa 2 O 5 Ta RRAM device. VLSI Technology (VLSI-Technology): Digest of Technical Papers, 2014 Symposium on; 2014: IEEE.

25.     Dan Y, Poo M-m. Spike timing-dependent plasticity of neural circuits. Neuron. 2004;44(1):23-30.

26.     Ban S, Kim O. Improvement of switching uniformity in HfOx-based resistive random access memory with a titanium film and effects of titanium on resistive switching behaviors. Japanese Journal of Applied Physics. 2014;53(6S):06JE15.

27.     Scopel W, da Silva AJ, Orellana W, Fazzio A. Comparative study of defect energetics in HfO2 and SiO2. arXiv preprint cond-mat/0310747. 2003.

28.     Hardacre C, Roe GM, Lambert RM. Structure, composition and thermal properties of cerium oxide films on platinum {111}. Surface science. 1995;326(1):1-10.

29.     Hasegawa T, Shahed SMF, Sainoo Y, Beniya A, Isomura N, Watanabe Y, et al. Epitaxial growth of CeO2 (111) film on Ru (0001): Scanning tunneling microscopy (STM) and x-ray photoemission spectroscopy (XPS) study. The Journal of chemical physics. 2014;140(4):044711.

30.     Larachi Fç, Pierre J, Adnot A, Bernis A. Ce 3d XPS study of composite CexMn 1− xO 2− y wet oxidation catalysts. Applied Surface Science. 2002;195(1):236-50.

31.     Guy J, Molas G, Blaise P, Carabasse C, Bernard M, Roule A, et al., editors. Experimental and theoretical understanding of forming, SET and RESET operations in conductive bridge RAM (CBRAM) for memory stack optimization. Electron Devices Meeting (IEDM), 2014 IEEE International; 2014: IEEE.

32.     Khafizov M, Park IW, Chernatynskiy A, He L, Lin J, Moore JJ, et al. Thermal conductivity in nanocrystalline ceria thin films. Journal of the American Ceramic Society. 2014;97(2):562-9.

33.     Menzel S. Modeling and simulation of resistive switching devices: Lehrstuhl für Werkstoffe der Elektrotechnik II und Institut für Werkstoffe der Elektrotechnik; 2013.

34.     Volkov YA, Palatnik L, Pugachev A. Investigation of the thermal properties of thin aluminum films. Zh Eksp Teor Fiz. 1976;70:2244-50.

35.     Panzer MA, Shandalov M, Rowlette J, Oshima Y, Chen YW, McIntyre PC, et al. Thermal properties of ultrathin hafnium oxide gate dielectric films. Electron Device Letters, IEEE. 2009;30(12):1269-71.

36.     Alibart F, Zamanidoost E, Strukov DB. Pattern classification by memristive crossbar circuits using ex situ and in situ training. Nature communications. 2013;4.

37.     Eryilmaz SB, Kuzum D, Jeyasingh R, Kim S, BrightSky M, Lam C, et al. Brain-like associative learning using a nanoscale non-volatile phase change synaptic device array. Frontiers in neuroscience. 2014;8.

38.     Wong H-SP, Lee H-Y, Yu S, Chen Y-S, Wu Y, Chen P-S, et al. Metal–oxide RRAM. Proceedings of the IEEE. 2012;100(6):1951-70.

39.     Rajendran B, Liu Y, Seo J-s, Gopalakrishnan K, Chang L, Friedman DJ, et al. Specifications of nanoscale devices and circuits for neuromorphic computational systems. Electron Devices, IEEE Transactions on. 2013;60(1):246-53.

# Supplementary Information

## A sub-1-volt analog metal oxide memristive-based synaptic device for energy-efficient spike-based computing systems

Cheng-Chih Hsieh[3], Anupam Roy[1], Yao-Feng Chang[1], Davood Shahrjerdi[4], and Sanjay K. Banerjee[1]

### 1. X-ray photoelectron spectroscopy (XPS) studies

We performed a series of material optimization and characterization experiments in order to engineer the structural properties of the $HfO_x$ capping layer using atomic layer deposition (ALD). The XPS measurements were used to study the chemical bonding and the stoichiometric ratio of the $CeO_x$ and $HfO_x$ films. All XPS spectra were acquired at room temperature using a Vacuum Generator Scientific SCALAB Mark II system and monochromatic Al $K_\alpha$ ($h\nu$= 1486.7eV) X-ray radiation source. The background pressure was kept below $7.5\times10^{-8}$ Torr during the XPS measurements. The XPS curves were calibrated to the carbon 1s peak. Previous XPS studies of $CeO_x$ suggest that the Ce 3d XPS core-level spectrum has three-lobed envelops (which are located at around 882-890 eV, 895-910 eV and 916 eV) that originate from the different final states of mixed valency [1-3]. In Ce 3d spectrum, the u''' (v '''), u'' (v'') and u (v) denote the $Ce^{4+}$ final states and correspond to Ce $3d^94f^0$ O $2p^6$, Ce $3d^94f^1$ O $2p^5$ and Ce $3d^94f^4$ O $2p^4$ respectively for Ce3d$_{3/2}$ and Ce3d$_{5/2}$. The $Ce^{3+}$ final states in Ce 3d spectrum correspond to $3d^94f^1$ O $2p^5$ and $3d^94f^2$ O $2p^4$ and are labeled as u' (v') and $u_o$ ($v_o$) in the XPS spectrum. Previous reports also suggest that the u' (v') peaks are typically stronger than the $u_o$ ($v_o$) peaks.

### 2. Accelerated retention test

Another important metric in memristors is device reliability in terms of endurance and retention. Accelerated retention test was performed at 150°C. The measurements were made at the constant stress voltage of +0.2V. To find the corresponding data retention time at the working temperature, we calculated the acceleration factor (AF) using the following equation [4]:

$$AF = \exp\left(\frac{E_a}{k} * \left(\frac{1}{T_{Work}} - \frac{1}{T_{acc}}\right)\right) \qquad (1)$$

where $E_a$ is the intrinsic activation energy, $k$ is the Boltzmann constant, $T_{Work}$ is the working temperature (here assumed to be 50°C), and $T_{acc}$ is the temperature at which the accelerated retention test was performed (that is 150°C). For the $CeO_x$-based memristors, $E_a$ was assumed to be 0.85eV, which is a conservative estimated value. Substituting the above values in equation (1), the acceleration factor was calculated to be about 1361. Since the devices in Figure 4 (See Main text) exhibit data retention in excess of 10⁵ seconds at 150°C, their expected retention time at the working temperature of 50°C easily exceed 10⁸ seconds, which corresponds to 10 years.

---

3 Microelectronics Research Center, The University of Texas at Austin, Austin, Texas 78758, USA.

4 Electrical and Computer Engineering, New York University, Brooklyn, New York, 11201, USA.

## 3. Generation of waveforms for Spike-Timing-Dependent Plasticity (STDP) measurements

Multiple square waves with various pulse heights were tailored for emulating exponential decay pulses used in STDP studies. The time interval of the pulses on the top and bottom electrodes (that are pre- and post-neural spikes) was varied relative to each other while monitoring the conductance of the device. The pulses were created using the Keysight B1500 pulse generator unit. Depending on the time difference between the pulses illustrated in Figure S1 (a) and (b), the device demonstrates long-term depression ($\Delta t<0$) and potentiation ($\Delta t<0$) (see Figure 4, main text).

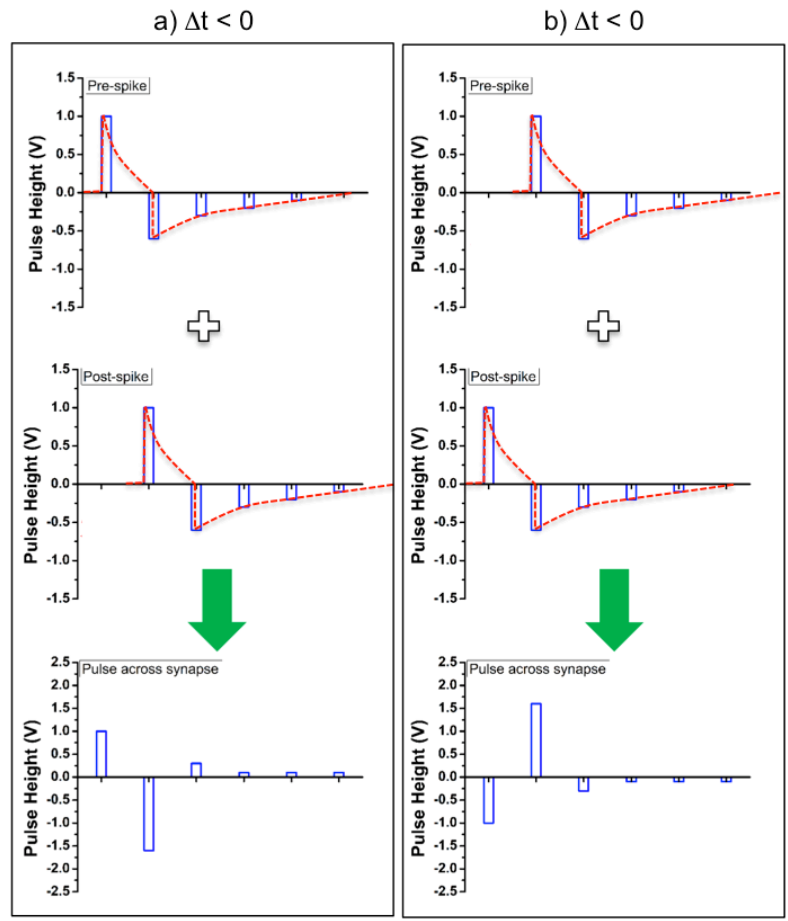

**Figure S1:** Schematic illustration of the pre- and post-neural spikes that were used for implementing the STDP learning in our bilayer memristive devices. A series of square pulses (solid line) were generated to emulate the exponential spikes (dashed line).

# References:

1.      Hardacre C, Roe GM, Lambert RM. Structure, composition and thermal properties of cerium oxide films on platinum {111}. Surface science. 1995;326(1):1-10.

2.      Hasegawa T, Shahed SMF, Sainoo Y, Beniya A, Isomura N, Watanabe Y, et al. Epitaxial growth of CeO2 (111) film on Ru (0001): Scanning tunneling microscopy (STM) and x-ray photoemission spectroscopy (XPS) study. The Journal of chemical physics. 2014;140(4):044711.

3.      Larachi Fç, Pierre J, Adnot A, Bernis A. Ce 3d XPS study of composite CexMn 1− xO 2− y wet oxidation catalysts. Applied Surface Science. 2002;195(1):236-50.

4. Martin, N., and K. Peter. "Typical data retention for nonvolatile memory." Engineering Bulletin of Freescale semiconductor, EB618/D, Rev 4 (2005).